# Percolation of Immobile Domains in Supercooled Thin Polymeric Films

Arlette Baljon, Joris Billen, Rajesh Khare
*Department of Physics, San Diego State University, San Diego, CA 92128*

We present an analysis of heterogeneous dynamics in molecular dynamics simulations of a thin polymeric film near the glass transition. The film is supported by an absorbing structured surface. It is simulated using a coarse-grained bead-spring model, along with a specific criterion to select slow - "immobile"- beads. As it turns out, the immobile beads occur throughout the film, yet their distribution is inhomogeneous, with the probability of their occurrence decreasing with larger distance from the substrate. Still, enough immobile beads are located near the free surface to cause them to percolate in the direction perpendicular to substrate surface, at a temperature near the glass transition temperature. This result is in agreement with a recent theoretical model of glass transition.

Ultrathin films of polymers have received ample attention in recent literature, due to their relevance for a wide range of applications. They are used as masks in lithographic processes in the microelectronics industry and their response to stress is relevant to the overall dynamical properties of polymer-nanoparticle composites [1]. Experiments [2] and simulations [3,4] alike have shown that, depending on polymer-surface interaction energy, the glass transition temperature ($T_g$) of such thin films can be either substantially higher or lower than the corresponding $T_g$ of the bulk polymers. Various theories, such as the layer model, have been proposed to explain this observed deviation of the $T_g$ from its bulk value [5].

Long and coworkers have recently proposed a phenomenological model based on thermally induced density fluctuations [6]. They hypothesize that the glass transition is governed by the percolation of small (size ~ 2 nm) domains of high density. They argue that, for a bulk system, the glass transition is governed by three-dimensional percolation of immobile domains. Since percolation in two dimensions requires a larger fraction of immobile domains than in three dimensions, the $T_g$ of a thin film repelled by the substrate is below that of the bulk value, a result in agreement with experimental observations. On the other hand, with an attractive surface, glass transition is governed by percolation of immobile domains in the direction perpendicular to the film plane. Hence, the $T_g$ of such a system is above that of the bulk value. Moreover, by assuming that slow domains correspond to high-density regions, the model explains the heterogeneous dynamics that is observed close to the glass transition in a number of experimental studies [7].

The notion that the observed change in mechanical properties at a glass transition could be caused by the percolation of domains of high density throughout the system has been around for a few decades but is still an unresolved issue. Dynamic heterogeneities in glass-forming systems have been widely discussed over the past decade. Whereas one might naively expect that mobility of particles or polymeric segments directly varies with the local density, recent studies [8,9,10] have shown that, in fact, it more directly correlates with the details of the local energy landscape. In this work, we investigate percolation properties of domains of low mobility.

Several recently published computational studies have focused on dynamic heterogeneities in glass forming systems [11,12,13]. Supercooled melts of non-entangled bead-spring-model polymer chains have been investigated, as well as Lennard-Jones mixtures of spherical particles. A few have dealt with heterogeneous dynamics in thin films [8,10]. The mobility of particles is commonly defined as their displacement during a certain time interval $\Delta t$. As will be explained in more detail below, the value of $\Delta t$ is chosen to maximize the dynamic heterogeneity. Mobility also has been defined as the maximum distance traveled by a particle from its initial position during $\Delta t$ [11]. At low values of mobility, the results appear to depend on the definition of mobility.

It has been found that "mobile" particles (defined as a fraction, typically 5% or 6.5%, of the beads exhibiting the highest displacement over $\Delta t$) form string-like clusters. It was also shown that the motion of particles in these strings is correlated [13] and that high particle mobility propagates through the system. These observations support the concept of dynamic facilitation [14]. In Lennard-Jones mixtures, the size of the "mobile" clusters diverges at the critical mode coupling theory (MCT) temperature [11]. However, in studies of polymeric melts [12], these clusters were found to be dynamic *i.e.* their size was found to increase or decrease with time. Although it was shown that "immobile" particles (defined as a certain small fraction, say 5%, of particles with the lowest displacement) tend to form compact clusters, the thrust of previous investigations has been the behavior of mobile particles. By contrast, in the present study, we focus on the immobile ones. Our work shows that, near the $T_g$, the immobile beads start percolating *i.e.* they extend across the thickness of the film, from the substrate surface to the free surface.

The simulations reported in this letter were carried out on an ultrathin film of a polymeric material, supported by a substrate surface. The other surface of the film is free. The polymers are modeled using a bead-spring model [15]. The film consists of 40 polymer chains, each containing 100 beads. Non-bonded interactions between the chain beads have been modeled using a standard Lennard-Jones (LJ) potential. All quantities are expressed in terms of the parameters ($\sigma,\varepsilon$) of this potential. Beads connected by the chain structure interact through the usual FENE potential. The substrate surface consists of an idealized FCC lattice with a spacing of 1.3. For the interaction between polymer beads and wall atoms, a LJ potential with $\sigma=0.8$ and $\varepsilon=1$ has been employed. The film thickness is about 13 (in reduced units), its in-plane dimensions are approximately 19 x 16. Periodic boundary conditions are employed in the plane of the film. Molecular dynamics simulations were carried out at constant temperature by coupling the system to a heat bath. Various temperatures spanning the glass transition were simulated. Initial states were prepared by cooling the system at a rate of 1/25,000 and subsequently equilibrating it for at least 100,000 $\tau$. More details on the simulation method can be found in our previous work [16].

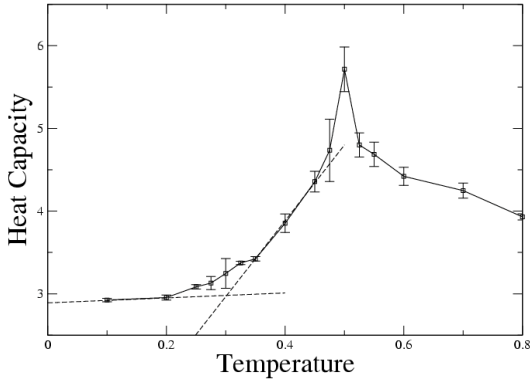

FIG. 1. Heat capacity (Cp) of the film as a function of temperature.

The $T_g$ is determined by measuring the heat capacity ($C_p$). Fig. 1 displays $C_p$ as a function of temperature. These data show a distinct peak at 0.5. The onset of the rise in $C_p$ is at 0.3 and corresponds to the crossover point of the two linear fits to the data as shown. This temperature has been called the fictive glass transition temperature [17]. In previous work [16], two other methods were also used to determine the $T_g$. First, film thickness versus temperature data show a crossover at $T_g = 0.51$. Second, $T_g$ was determined by the divergence of the relaxation time obtained from bead diffusion (MCT critical temperature), and was found to be 0.36. In the present study, emphasis is on determining the molecular changes exhibited by the film at the glass transition. To this end, we have determined the "immobile" beads in the system by considering the radius of gyration of the path traveled by a bead during $\Delta t$. Although this definition provides a more comprehensive representation of the particle mobility based on its trajectory in a given time, it is more computationally intensive than the definitions of bead mobility used in previous work [11]. A histogram of bead mobility values obtained using $\Delta t = 5$ is displayed in Fig. 2 for two temperatures.

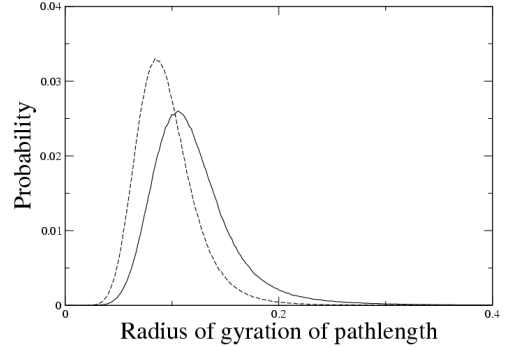

FIG. 2. Histogram of radius of gyration of path length ($R_{gp}$) for $\Delta t = 5$. Data are shown for T = 0.25 (long dashed) and T = 0.35 (solid). The vertical axis gives the probability that a bead has an $R_{gp}$ within a 0.002 interval.

The choice of the particular value for $\Delta t$ deserves some discussion. As found in both experiments and simulations, the heterogeneity as exhibited by the bead diffusion peaks at a specific time interval, of the order of the stress relaxation time [10]. This time interval can be determined by considering the first non-Gaussian correction to the incoherent dynamic structure factor $\alpha_2$. When we measured this interval, we found that it not only depends on temperature, but also on the distance from the substrate. Smith et al. [10] have discussed this issue in detail in a recent letter. At T=0.35 we find that, averaged over the entire film, $\Delta t=200$. In experiments on heterogeneous dynamics in colloidal suspensions, Weeks et al. [18] have designated the relaxation time corresponding to the structural alpha relaxation as $\Delta t$. For the thin film under investigation, we have reported these relaxation times in our previous work [16]. They range from 2 at T = 0.6 to 50 at T = 1.2. In light of these observations, we have performed our calculations using three different values of $\Delta t = 5$, 10 and 20.

Given a histogram of the radius of gyration of bead path lengths, the next task is to define a specific criterion to characterize the beads as immobile. If a fixed fraction, *e.g.* 5%, of the beads with the smallest value of radius of gyration of path length ($R_{gp}$) are defined as immobile, then the definition of immobility (range of values of $R_{gp}$) depends on temperature. With this definition, the size of the clusters of immobile particles is relatively independent of the temperature [11]. In this study, the Lindemann criterion for melting [19] is used to define a threshold value of $R_{gp}$ below which the beads are considered immobile. Simple solids are known to melt when the average distance by which the atoms fluctuate around their equilibrium positions is more than 0.1-0.15 in units of the lattice constant. For irregular systems this critical value turns out to be slightly lower [19]. Given this, beads with a radius of gyration of path length below 0.1 σ are considered immobile in this work.

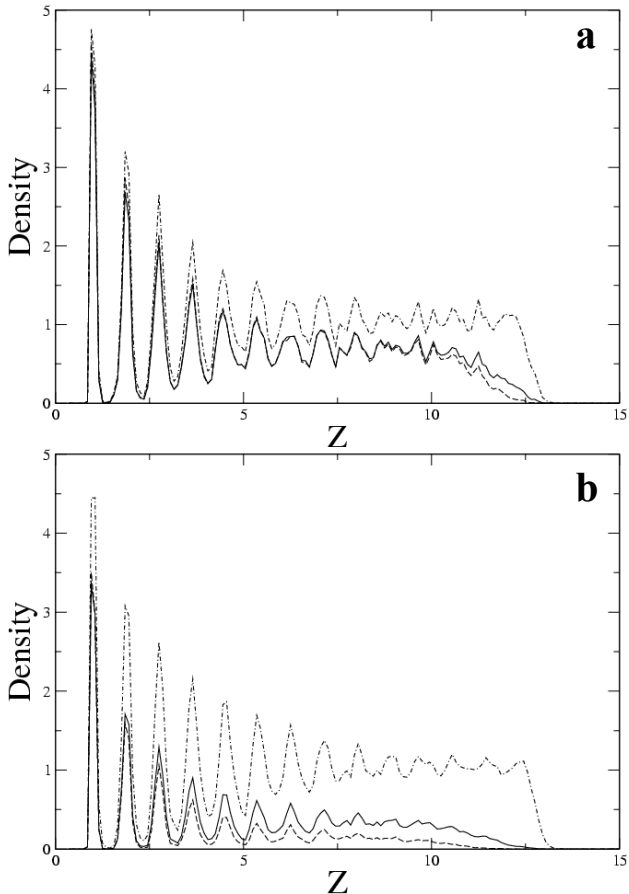

FIG. 3. Number density of the immobile particles as a function of distance from the substrate. Immobile particles are selected either using Δt=5 (solid) or Δt=20 (long dashed). Data are compared with the number density of all beads (dot-dashed). Two temperatures (a) T = 0.25 and (b) T = 0.35 are shown.

Fig. 3 compares the distribution of the selected immobile beads with the overall bead distribution in the film. The observed layering in the plot of the density as a function of distance from the substrate is typical for thin supported films [4]. Fig. 3 shows that the immobile beads occur in each layer, but their concentration decreases with distance from the substrate surface. Previous simulations have calculated average values for the bead mobility in each layer and concluded that they are a function of the distance from the substrate [4,20]. To our knowledge, this is the first study that focuses on the distribution of immobile beads in the film. A significant finding is that these immobile beads are distributed over the entire film and that even layers far away from the substrate contain a few of them. If all of the immobile beads resided in layers near the substrate, they would not be able to percolate across the thickness of the film.

Once immobile beads in the system are selected, a cluster analysis is performed on them to determine if they percolate across the thickness of the film. Four hundred sets of immobile beads, selected at different time instants during the simulated trajectory, are analyzed for different values of temperature and Δt. Immobile beads are considered to belong to the same cluster if they are situated at a distance of less than 1.45 from each other (where 1.45 equals the distance at which the first minimum in the radial distribution function is observed). A cluster thus defined is considered to percolate if at least one bead in the cluster is in the layer next to the substrate, and another in the layer at the free surface. Fig. 4 displays the fraction of these 400 sets that contain a percolating cluster. We find that this percolation probability changes significantly over a small temperature range. It is also interesting to note that the percolation probability is not merely a function of the fraction of immobile beads, as in the numerical examples by Long et al. [6]. This is due to the fact that the immobile beads are not randomly distributed over the film. For instance, we found that, by lowering the temperature and decreasing Δt, a selection of immobile beads of the same size can be obtained. However, this selection contains a larger number of immobile beads near the free surface and hence is more likely to percolate. We have found that the percolation transition is very sensitive to the properties of the layer near the free surface. When an appropriate fraction of the beads in this layer is dynamically arrested by the surrounding ones during Δt, percolation becomes likely.

In the above analysis, percolation is studied by focusing on regions of low mobility. An interesting issue to investigate is whether the regions of low mobility also correspond to the regions of high local density. A detailed investigation of the local density values is beyond the scope of this work. However, we have compared the local environment of an "immobile" bead with that of an "average" bead by computing the radial distribution function of both types. The functions obtained from a weighted average of the two-dimensional distribution function of each layer, are shown in Fig. 4. The peaks are slightly higher for the "immobile" beads, consistent with the notion that the local density and the potential energy landscape are different around the immobile beads.

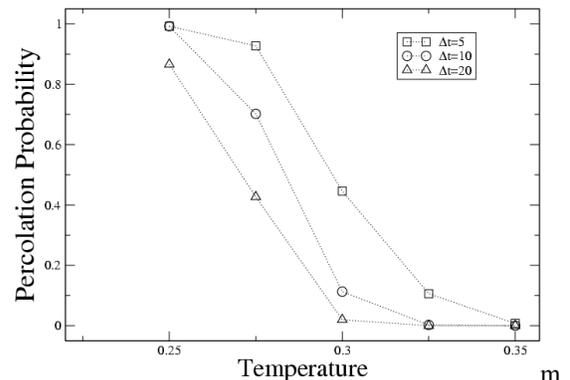

FIG. 4. Probability that the set of immobile particles percolates for several values of Δt and temperature.

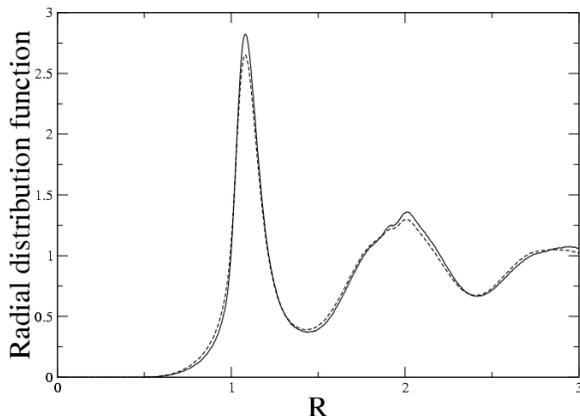

FIG. 5. A comparison of the radial distribution functions of all of the beads (solid line) and immobile beads (dashed line) at T = 0.3 and $\Delta t$ = 5.

In summary, we have shown that, in simulated thin polymer films supported by an attractive surface, clusters of immobile beads start to percolate in the direction perpendicular to the substrate at temperatures just below the glass transition. Furthermore, this "percolation transition" is found to occur over a small temperature range. The actual value of the critical temperature for which there is a 50% probability that a cluster of immobile beads will percolate depends on $\Delta t$, the time-interval used for selecting the immobile beads. For $\Delta t$ = 5, the critical temperature has a value of 0.3, which coincides with the temperature at which heat capacity begins to rise above it's value in the glassy state. Our data show that immobile beads are distributed throughout the film, but show a preference for the region near the attracting substrate. They favor regions with high-energy barriers for bead motion. The $T_g$ of polymer films is known to show a strong dependence on the details of the polymer-substrate interaction and the film thickness. Thus, an interesting issue to investigate in future work will be whether the critical temperature for percolation shows a dependence on these factors in a similar fashion.

The authors gratefully acknowledge support by a grant from the Donors of the Petroleum Research Fund. Moreover, we acknowledge Prof M.A.J. Michels for bringing the theoretical percolation model to our attention. JB contributed to this work during a study-abroad internship and acknowledges the Department of Applied Physics at the Technical University of Eindhoven for financial support.